\documentstyle[epsfig,twocolumn,pra,aps]{revtex}
\begin{document}
\draft

\wideabs{

  \title{Entanglement transfer from continuous variables to qubits}

   \author{W. Son,$^{1,2,3}$ M. S. Kim,$^1$ Jinhyoung Lee,$^3$ and D. Ahn,$^3$}

  \address{$^1$  School of Mathematics and Physics, The Queen's University,
    Belfast BT7 1NN, United Kingdom\\
    $^2$ Department of Physics, Sogang University,
    CPO Box 1142, Seoul 100-611, Korea \\
    $^3$ Institute of Quantum Information Processing and
    Systems, University of Seoul, Tongdaemoon-ku, Seoul,
    Korea}

  \date{\today}

  \maketitle

\begin{abstract}
We show that two qubits can be entangled by local interactions
with an entangled two-mode continuous variable state. This is
illustrated by the evolution of two two-level atoms interacting
with a two-mode squeezed state.  Two modes of the squeezed field
are injected respectively into two spatially separate cavities and
the atoms are then sent into the cavities to resonantly interact
with the cavity field.  We find that the atoms may be entangled
even by a two-mode squeezed state which has been decohered while
penetrating into the cavity.
\end{abstract}

\pacs{PACS number(s);42.50.Dv, 03.65.Bz, 42.50.-p}

}

\section{\bf Introduction}

Entanglement is one of the important ingredients in current
development of quantum information processing.  If the density
operator of a bipartite system, $\hat\rho$, is not represented by
a convex sum of product states, {\i.e} $\hat\rho\neq\sum_i
\hat\rho_a (i)\otimes\hat\rho_b(i)$ where $\hat\rho_a$ and
$\hat\rho_b$ are density operators for subsystems $a$ and $b$, the
system is said to be entangled. For a $2\times2$-dimensional
space, Peres and Horodecki {et
  al.}\cite{Peres} found a criterion of entanglement: When the
partial transposition of its density matrix has a negative
eigenvalue, the bipartite system is entangled.  Entanglement is a
purely quantum nature and it sometimes presents a contrast to
nonlocality (see \cite{Bennett-many} for further discussions).
Even though the two concepts coincide for pure
bipartite states \cite{Gisin}, when mixed states are concerned,
there may be mixed entangled states which do not show nonlocality \cite{Werner}.

In this paper, we are interested in a possibility of entangling two
remote qubits using a bipartite continuous variable state.  This is
illustrated by the evolution of two two-level atoms interacting with a
two-mode squeezed state as shown in Fig. 1.  Assume that two modes of
the squeezed field are injected respectively into two spatially
separate cavities.  The atoms are then sent into the cavities to
resonantly interact with the cavity field. A high-Q cavity is assumed
to consider the Jaynes-Cummings type interaction between the atom and
cavity field \cite{Jaynes-Cummings}.

A pair of atoms can be prepared in an entangled state using the
atom-field interaction in a high-$Q$ cavity. The interaction of a
single two-level atom with a cavity field brings about
entanglement of the atom and the cavity field \cite{Phoenix93}. If
the atom does not decay into other internal states after it comes
out from the cavity, the entanglement will survive and it can be
transferred to a second atom interacting with the cavity field.
The violation of Bell's inequality can be tested by the local
measurement of atomic states.

There are proposals to entangle fields in two spatially separated
cavities using the atom-field interaction
\cite{Meystre92,Davidovich93}.  A two-level atom in its excited
state passes sequentially through two resonant single-mode vacuum
cavities and is found to be in its ground state after the
second-cavity interaction.  If the interaction with the first
cavity is equivalent to a $\pi/2$ vacuum pulse and the
second-cavity interaction is to a $\pi$ pulse  then the atom could
have deposited a photon either in the first cavity or in the
second so that the final state $|\Psi_f\rangle$ of the two cavity
field is \cite{Meystre92}
\begin{equation}
\label{Meystre}
|\Psi_f\rangle = {1 \over \sqrt{2}}(|1,0\rangle +
\mbox{e}^{i\varphi}|0,1\rangle)
\end{equation}
where $|1,0\rangle$ denotes one photon in the first cavity and
none in the second, and $|0,1\rangle$ vice versa and the
nonlocality of the cavity fields can also be tested using local
parity measurement \cite{Kim00}. Using the entangled cavity field
(\ref{Meystre}), an unknown atomic quantum state can also be
teleported \cite{David}.

In quantum algorithm, the entangled state (\ref{Meystre}) of
$\phi=0$ is generated by the action of a controlled-NOT gate if
the two input particles are in
$|0\rangle(|0\rangle+|1\rangle)/\sqrt{2}$.  Type-II \cite{Kwiat1}
or type-I \cite{Kwiat2} parametric down-conversion may produce
entangled states. By putting one photon state into a beam
splitter, an entangled state is generated in the two output ports
\cite{Tan}.  The beam splitter may also be used to produce a continuous-variable
entangled state \cite{Kim01}. By a non-degenerate optical
parametric amplifier a two-mode squeezed state, which is a
continuous-variable entangled state, is generated
\cite{Braunstein}.

We show the initial preparation of the cavity in Section II and
the evolution of atoms and field in Section III.  The evolution of
entanglement between two atoms is studied in Section IV.

\section{Cavity field}

The two-mode squeezed state \cite{Caves}, which is generated using
a non-degenerate optical parametric amplifier, is an entangled
continuous-variable state. Maximal entanglement is not possible to
produce for two-mode squeezed states because such a state implies
infinite energy. The action of the two-mode squeeze operator,
$\hat S(s)= \exp(-s\hat{a}\hat{b}+s\hat{a}^\dag\hat{b}^\dag)$ on
the vacuum $|0,0\rangle$ produces the two-mode squeezed state
\begin{equation}
\label{tmsqueeze} |\psi_F\rangle = \frac{1}{\cosh s}
\sum^{\infty}_{n=0} (\tanh s)^n |n,n\rangle
\end{equation}
where $s$ is the squeezing parameter. Here $\hat a$ and $\hat b$ ($\hat a^\dag$ and $\hat b^\dag$)
are bosonic annihilation (creation) operators.
The amount of entanglement of the two-mode squeezed state
is linearly proportional to the squeezing parameter\cite{vanEnk}.

Each mode is injected to the cavity which is initially in the vacuum state as shown in Fig. 1.
A complete quantum-mechanical picture of injecting an external
field to a high-Q cavity is gained by modelling this operation
with beam splitters \cite{Kim98}.  For example, injecting
$\hat\rho_f$ into a cavity initially in the vacuum, the cavity
field becomes
\begin{equation}
\label{coupling}
\hat\rho_c=\mbox{Tr}_f\hat{B}(\theta)\hat\rho_f|0\rangle_c\langle
0|\hat{B}^\dag(\theta)
\end{equation}
where $\mbox{Tr}_f$ is tracing over the mode $f$ and the coupling
between the external field and cavity is determined by the beam
splitter operator
$\hat{B}(\theta)=\exp[\frac{\theta}{2}(\hat{c}\hat{f}^{\dagger}-\hat{c}^{\dagger}\hat{f})]$
with $\hat{c}$ and $\hat{f}$ being annihilation operators
respectively for the cavity and external fields and $\hat{c}^\dag$
and $\hat{f}^\dag$ are their hermitian conjugates. The reflection
coefficient $r=\cos(\theta/2)$ of the beam splitter is determined
by the coupling between the external field and cavity.

Extending Eq.(\ref{coupling}) to coupling of the two-mode squeezed
state (\ref{tmsqueeze}) with two independent cavities initially in
their vacuum states, we obtain the density operator for the cavity
field, after injecting the squeezed state:
\begin{eqnarray}
\label{mixed-field} \hat{\rho}_{c} &=& \left(\frac{1}{\cosh
s}\right)^2 \sum^{\infty}_{n,m=0}
\sum^{\min[n,m]}_{k,l=0} (\tanh s)^{n+m} G^{nm}_{kl}(\theta)\\
&&~~~~~~~~~~~~\times |n-k,n-l\rangle\langle m-k,m-l|\nonumber
\end{eqnarray}
where $G^{nm}_{kl}(\theta)=
C^{n}_{k}(\theta)C^{m}_{k}(\theta)C^{n}_{l}(\theta)C^{m}_{l}(\theta)$
with
\begin{equation}
C^{n}_{k}(\theta) = \sqrt{\frac{n!}{k!(n-k)!}}\cos^k{\theta \over 2}
\sin^{n-k}{\theta \over 2}\nonumber.
\end{equation}
The density matrix $\hat{\rho}_{c}$ represents the composite
system of the cavity fields in a mixed state.  It has been shown
that the mixed state $\hat\rho_c$ in Eq.(\ref{mixed-field}) is
never separable regardless of $\theta$ using the separability
criterion for a Gaussian field \cite{Kim+Lee,Duan}.

\section{Atom-field interaction}
The Jaynes-Cummings model \cite{Jaynes-Cummings} is one of the
most studied models in foundations of quantum mechanics.  It
consists of a two-level atom resonantly coupled to a single-mode
radiation field inside a cavity.  The importance of the
Jaynes-Cummings model is many-fold. This exactly soluble simple
model may serve a basis for more complicated systems and it
manifests interesting quantum nature of the field and atom.  This
model can also be experimentally realisable \cite{Haroche}.

The Hamiltonian for the Jaynes-Cummings model in the interaction
picture is written in atomic bases $\{|e\rangle,|g\rangle\}$
\cite{Jaynes-Cummings}
\begin{equation}
\hat{H}_I=\pmatrix{ 0&\lambda\hat{c}\cr \lambda\hat{c}^{\dagger}&0}
\end{equation}
where $\lambda$ is the coupling coefficient between the atom and
field. The corresponding unitary evolution operator
$\hat{U}=\exp(-i\hat{H}_{I} t)$ can be written in the following
form\cite{stenholm}.
\begin{equation}
\hat{U}(t)=\pmatrix{\cos\hat{\Omega}_{n+1}t &
-i\lambda\hat{a}\frac{\sin\hat{\Omega}_n t}{\hat{\Omega}_n}\cr
 -i\lambda\hat{a}^{\dagger}\frac{\sin\hat{\Omega}_{n+1}
t}{\hat{\Omega}_{n+1}} & \cos\hat{\Omega}_n t }
\end{equation}
where the Rabi-frequency operator, $
\hat{\Omega}_{n}=\lambda\sqrt{\hat{a}^\dag\hat{a}}$.

In our model, there are two atoms residing in two separate
cavities. The total unitary operator, thus, can be constructed by
a direct product of two unitary operators representing evolutions
of two independent atom-field interactions:
$\hat{U}_T(t)=\hat{U}^A(t)\otimes\hat{U}^B(t)$ where $A$ and $B$
are the labels for two cavities. For simplicity, we assume that
the atoms undergo the same evolution with the same atom-field
coupling constants and interaction times.

\section{Entanglement of two atoms}
We assume that the two atoms are prepared in a separable
pure state $\hat{\rho}_a$ and the field is in the mixed
two-mode squeezed state (\ref{mixed-field}).  The composite system of
the atoms and field evolves unitarily and is
represented by the density operator $\hat{\rho}(t)$ at time $t$:
\begin{equation}
\hat{\rho}(t)=\hat{U}_T(t)~\hat{\rho}_a\otimes\hat\rho_c~
\hat{U}_{T}^\dag(t).
\end{equation}
Because we are interested in entanglement between the atoms, we
trace over the field variables and find the time-dependent density
operator $\hat{\rho}_a(t)$ for the two atoms:
\begin{equation}
\label{density}
\hat{\rho}_{a}(t)={\bf Tr}_{c}\{
\hat{\rho}(t)\}=\pmatrix{A(t)&0&0&E(t)\cr 0&B(t)&0&0\cr
0&0&C(t)&0\cr E(t)&0&0&D(t)}
\end{equation}
where the matrix basis is chosen as
$\{|e,e\rangle,|e,g\rangle,|g,e\rangle,|g,g\rangle\}$ and
$A(t),B(t),C(t),D(t)$, and $E(t)$ are the functions of interaction time,
which are determined by the initial state of the atoms. For
example, if the atoms are initially in $|g,g\rangle$ the functions
are given by
\begin{eqnarray}
\label{para} A(t)&=& \sum^{\infty}_{n=0} \sum^{n}_{k,l=0}
K^{nn}_{kl}(\theta,s)\sin^2(\lambda
t\sqrt{n-k}) \sin^2(\lambda t\sqrt{n-l})\nonumber\\
B(t)&=& \sum^{\infty}_{n=0} \sum^{n}_{k,l=0}
K^{nn}_{kl}(\theta,s)\sin^2(\lambda
t\sqrt{n-k}) \cos^2(\lambda t\sqrt{n-l})\nonumber\\
C(t)&=&\sum^{\infty}_{n=0} \sum^{n}_{k,l=0}
K^{nn}_{kl}(\theta,s)\cos^2(\lambda
t\sqrt{n-k}) \sin^2(\lambda t\sqrt{n-l})\nonumber \\
D(t)&=& \sum^{\infty}_{n=0} \sum^{n}_{k,l=0}
K^{nn}_{kl}(\theta,s)\cos^2(\lambda
t\sqrt{n-k}) \cos^2(\lambda t\sqrt{n-l})\nonumber\\
E(t)&=& \sum^{\infty}_{n=0} \sum^{n}_{k,l=0} K^{n+1
n}_{kl}(\theta,s) \sin(\lambda t\sqrt{n-k+1})\cos(\lambda
t\sqrt{n-k})
\nonumber\\
&&~~~~~~~~~~~~\times \sin(\lambda
t\sqrt{n-l+1}) \cos(\lambda t\sqrt{n-l})
\end{eqnarray}
where
\begin{equation}
\label{weight}
K^{nm}_{kl}(\theta,s)=\frac{(\tanh s)^{n+m}}{(\cosh
s)^2}~~ G^{nm}_{kl}(\theta).
\end{equation}
For the case that the atoms are initially prepared in
$|e,e\rangle$ the functions are modified from Eqs.(\ref{para}). In
this case, the functions can be obtained by switching
$\sin\leftrightarrow\cos$ and $k,l\rightarrow k+1,l+1$ in
Eqs.(\ref{para}).

The density operator $\hat{\rho}_a(t)$ characterizes the atoms
inside the cavity not in a pure state for $t>0$. There are several
ways to quantify a degree of entanglement for a mixed state such
as the quantum relative entropy\cite{vedral}, the entanglement
formation and the measure defined by negative eigenvalues for the
partial transposition of the density operator \cite{jinhyoung}.
Among the possible entanglement measure for a mixed state we take
the measure of negative eigenvalues for the partial transposition
of the density operator because the calculation is simple.

The partial transposition of $\hat{\rho}_a(t)$ has four
eigenvalues one of which may be negative.  The entanglement
measure ${\mathcal E}(\hat{\rho})$ is defined as multiplying the factor $-2$ to the sum of
negative eigenvalues. The entanglement measure then ensures the
scale between $0$ and $1$ and monotonously increases as
entanglement grows. The entanglement measure for the atoms inside
the cavities are given by the function of interaction time as
\begin{equation}
\label{entangle}
{\mathcal E}[\hat{\rho}_a(t)]=
\sqrt{[B(t)-C(t)]^2 +4E(t)^2 }-B(t)-C(t).
\end{equation}

The measure of entanglement in Eq.(\ref{entangle}) is not only the
function of interaction time but also the function of squeezing
parameter $s$.  It is obvious that entanglement does not
appear when the squeezing parameter $s= 0$ because the two-mode
squeezed state becomes separable and entanglement is not generated
only by local operations. For $s\neq 0$, we found that entanglement is
produced between the atoms inside the cavities regardless of $s$
when the atoms are initially prepared in $|g,g\rangle$ and $r=0$.

In Fig.\ref{fig:squeezing para}, entanglement between two atoms is
plotted against the squeezing parameter $s$ at a specific
interaction time $\lambda t=11$. (We have chosen this time because
the evolution of entanglement shows local maximum at this time as shown
later in Fig.~3.)   The figure is presented for the three
different values of the reflection coefficient $r$. It shows that the amount of
entanglement between the two atoms (qubits) is not a simple increasing
function of $s$ and even decreases as $s$ gets larger after it peaks at
$s\approx 0.65$.

The two-mode squeezed state is represented as a superposition of
the number states $|n,n\rangle$ as shown in Eq.(\ref{tmsqueeze}). The two
qubits which are in the ${\cal H}_{2}\otimes {\cal H}_{2}$ are
coupled with $|n,n\rangle$, $|n+1,n\rangle$, $|n,n+1\rangle$ and
$|n+1,n+1\rangle$ of continuous variables at the Rabi frequency
$\Omega_n=\lambda\sqrt{n}$. Because the amplitude $\tanh{s}$ in
Eq.(\ref{tmsqueeze}) increases with the squeezing parameter $s$,
the terms of higher excitation, {\em i.e.} large $n$, contribute
more to the atom-field interaction. The entanglement transfer is
then cancelled and approaches to zero for $s\rightarrow\infty$. It
means the maximally entangled continuous variable system does not
generate maximally entangled qubits by the unitary transformation we
have in this paper.

In Fig.\ref{fig:entanglement}, we plot the time evolution of
entanglement for $s=0.65$ for the different values of the
reflection coefficient $r$ when the atoms are initially prepared
in $|e,e\rangle$ and in $|g,g\rangle$. The amount of entanglement
decreases as the reflection coefficient $r$ is increased
while its dynamics exhibits the same qualitative
behaviors. The
reflection coefficient $r$ represents the degree of mixedness of
the two-mode squeezed state.  As $r$ gets larger, the amplitude
of the cavity field gets smaller.  However, it is interesting to
note that when atoms are initially in $|g,g\rangle$, even with $r$
as large as 0.99, the atoms may be entangled. Entanglement is
maximised when the cavities are initially prepared with a pure
two-mode squeezed state ($r=0$).

It is seen that entanglement evolves depending on the initial
preparation of atoms.  When the atoms are initially in
$|e,e\rangle$, entanglement does not appear at the early stage
of the interaction and undergoes fast oscillations. Compared with
the $|e,e\rangle$ preparation, the initial $|g,g\rangle$ state
produces more entanglement from the early stage of interaction.
Because of extra photons which are
emitted by atoms initially in $|e,e\rangle$, the atom-field interaction
becomes more complicated and brings about faster oscillations in
entanglement.

\section{Remarks}
We consider the entanglement of two two-level atoms by local
interaction with the two-mode squeezed field.  We found that the
entanglement is maximised when the two-mode squeezed field
is pure.  As far as the two-mode squeezed state is pure, the atoms
show entanglement during their evolution but the entanglement goes
to zero as $s\rightarrow\infty$. The dynamics of entanglement does not
change its time dependence while the overall amplitude is lowered when the initial squeezed
field becomes mixed.  We found that the
entanglement depends on initial preparation of atoms.

\begin{acknowledgements}
We thank the UK Engineering and Physical Sciences Research Council
for financial support through GR/R33304 and the Korean Ministry of
Science and Technology through the Creative Research Initiatives
Program under Contract No. 00-C-CT-01-C-35.  W.S. acknowledges the
British Council Scholarship and the BK 21 grant of the Korean Ministry of
Education (D-0099).

\end{acknowledgements}

\vspace{1.5cm}

\begin{figure}
  \begin{center}
\centerline{\scalebox{.5}{
\includegraphics{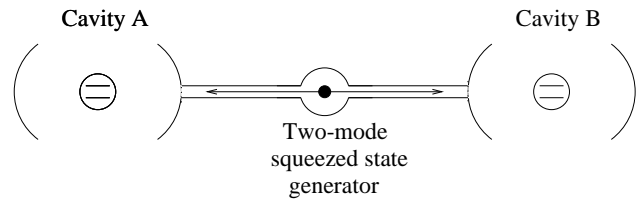}}}
\vspace{1cm}
    \caption{Each mode of the two mode squeezed state is injected into the high Q cavities A and B}
    \label{cavity}
  \end{center}
\end{figure}
\begin{figure}
  \begin{center}
\centerline{\scalebox{.25}{
\includegraphics{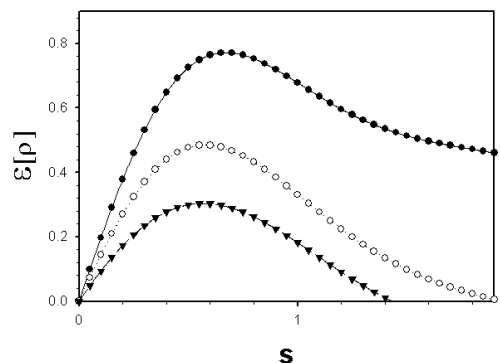}}}
    \caption{Entanglement between two atoms versus squeezing parameter $s$ at the interaction
    time $\lambda t = 11$.  The cavity fields are initially in the
pure state $r=0$ (solid circle), mixed state $r=0.25$ (open circle) and $r=0.7$
(triangle). The atoms are initially in $|g,g\rangle$.}
    \label{fig:squeezing para}
  \end{center}
\end{figure}
\begin{figure}
  \begin{center}
\centerline{\scalebox{.25}{
\includegraphics{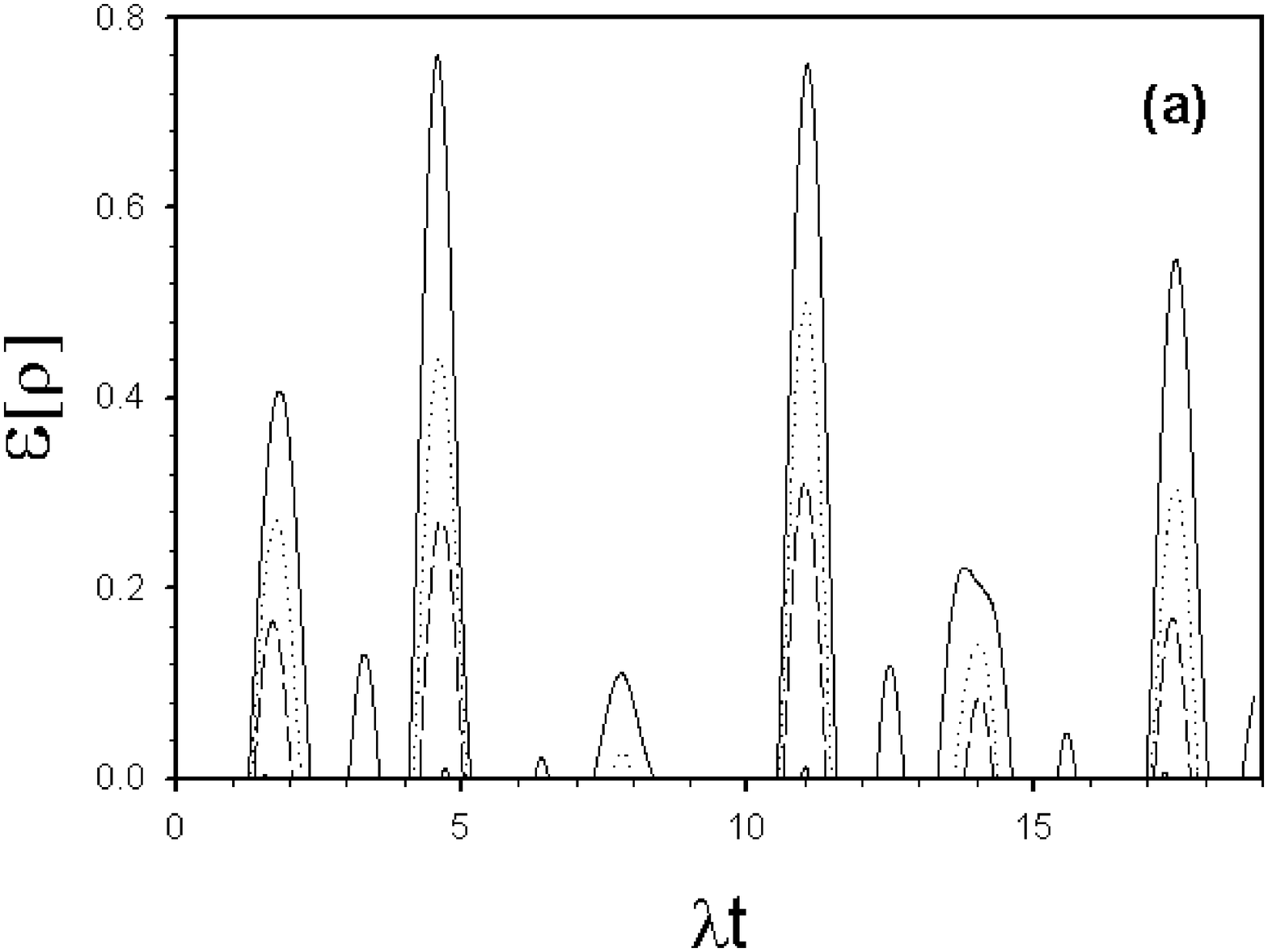}}}
\centerline{\scalebox{.25}{
\includegraphics{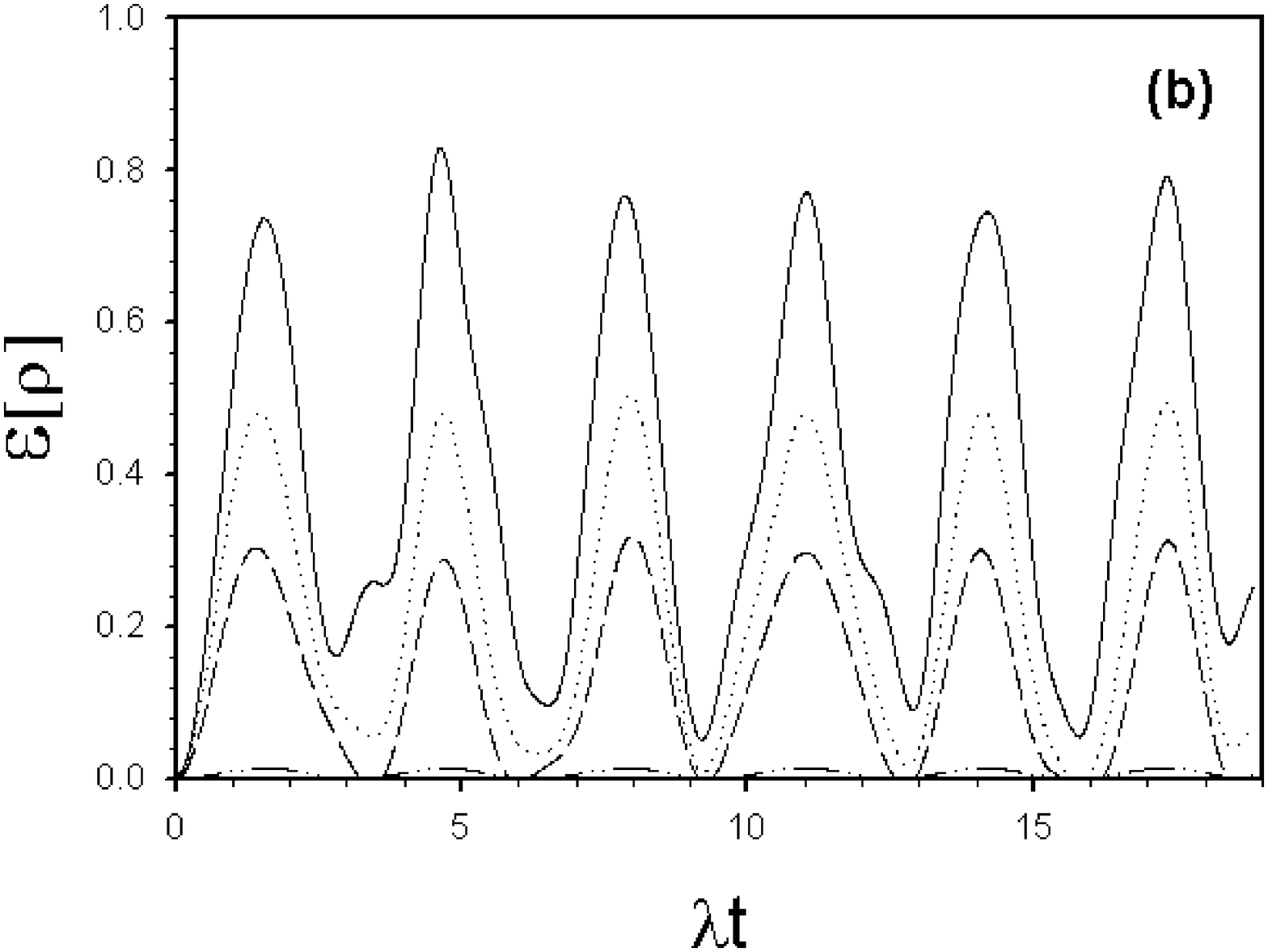}}}
    \caption{The time evolution of entanglement between the two atoms is plotted as the function of
dimensionless time $\lambda t$.  The cavity reflection parameters
are chosen as $r=0$ (solid), $r=0.25$ (dotted), $r=0.7$ (dashed)
and $r=0.99$ (dashed-dot).  The squeezing parameter is $s=0.65$.
Initial atomic states are prepared in $|e,e\rangle$ (a) and
$|g,g\rangle$ (b).}
    \label{fig:entanglement}
  \end{center}
\end{figure}


\begin{references}
\bibitem{Peres} A. Peres, \prl {\bf 77}, 1413 (1996); M. Horodecki, P.
  Horodecki, and R. Horodecki, Phys. Lett. A{\bf 223}, 1(1996).

\bibitem{Bennett-many} C.H. Bennett, D. P. DiVincenzo, C. A. Fuchs, T. Mor,
P. W. Shor, J. A. Smolin and W. Wootters, \pra {\bf 59}, 1070 (1999).

\bibitem{Gisin} N. Gisin, Phys. Lett. A {\bf 154}, 201 (1991).

\bibitem{Werner} R. F. Werner, \pra {\bf 40}, 4277 (1989).

\bibitem{Jaynes-Cummings} E. T. Jaynes and F. W. Cummings, Proc.
IEEE {\bf 51}, 89 (1963); B. W. Shore and P. L. Knight, \jmo {\bf
40}, 1195 (1993).

\bibitem{Phoenix93} S. J. D. Phoenix and S. M. Barnett, \jmo
{\bf 40}, 979 (1993); I. K. Kudryavtsev and P. L. Knight, \jmo
{\bf 39}, 1411 (1992).

\bibitem {Meystre92} P. Meystre, in {\it Progess in Optics XXX},
  edited by E. Wolf (Elsevier, Amsterdam, 1992).

\bibitem {Davidovich93} L. Davidovich, A. Maali, M. Brune, J. M. Raimond,
  and S. Haroche, \prl {\bf 71}, 2360 (1993).

\bibitem{Kim00} M. S. Kim and J. Lee, \pra {\bf 61}, 042102 (2000).

\bibitem{David} L. Davidovich, N. Zagury, M. Brune, J. M. Raimond, and S. Haroche, \pra {\bf 50}, 895 (1994).

\bibitem{Kwiat1} K. Mattle, H. Weinfurter, P. G. Kwiat and A. Zeilinger, \prl {\bf 76}, 4656 (1996).

\bibitem{Kwiat2} A. G. White, D. F. V. James, P. H. Eberhard and P. G. Kwiat, \prl {\bf 83}, 3103 (1999).

\bibitem{Tan} S. M. Tan, D. F. Walls and M. J. Collett, \prl {\bf 66}, 252 (1991).

\bibitem{Kim01} M. S. Kim, W. Son, V. Bu\v zek and P. L. Knight, LANL e-print, quant-ph/010636 (2001);
Ch. Silberhorn, P. K. Lam, O. Wei\ss, F. K\"onig, N. Korolkova and G. Leuchs, \prl {\bf 86},
4267 (2001).

\bibitem{Braunstein} S. Braunstein and H. J. Kimble, \prl {\bf 80}, 869 (1998);
A. Furusawa, J. L. Serensen, S. L. Braunstein, C. A. Fuchs, H. J.
Kimble, E.S. Polzik, Science {\bf 282}, 706 (1999).

\bibitem{vanEnk} S. Parker, S. Bose and M. Plenio, \pra {\bf 61},
032305 (2000).

\bibitem{Caves} B. L. Schumaker and C. M. Caves, \pra {\bf 31}, 3093
  (1985); S. M. Barnettand P. L. Knight, \jmo {\bf 34}, 841 (1987).

\bibitem{Kim98} M. S. Kim, G. Antesberger, C. T. Bodendorf and H.
Walther, \pra {\bf58}, R65 (1998).

\bibitem{Kim+Lee} J. Lee, H. Jeong and M. S. Kim, \pra {\bf 62},
032305 (2000).

\bibitem{Duan} L.-M. Duan, G. Giedke, J. I. Cirac and P. Zoller,
\prl {\bf 84}, 2722 (2000); R. Simon, {\em ibid}, 2726 (2000).

\bibitem{Haroche} J. M. Raimond, M. Brune and S. Haroche, Rev.
Mod. Phys. {\bf 73}, 565 (2001).

\bibitem{stenholm} S. Stenholm, Phys. Rep. {\bf 6C}, 1 (1973).

\bibitem{vedral} V. Vedral, M. B. Plenio, M. A. Rippin and P. L.
Kinght, \prl {\bf 78}, 2275 (1997).

\bibitem{jinhyoung} J. Lee, M. S. Kim, Y. J. Park and S.  Lee, \jmo
  {\bf 47}, 2151 (2000); J. Lee and M. S. Kim, \prl {\bf
    84}, 4236 (2000).

\end{references}
\end{document}